\documentclass[pre,floatfix,superscriptaddress,longbibliography,twocolumn,final,nofootinbib]{revtex4-1}
\usepackage{amsmath}
\usepackage{amssymb}
\usepackage{amsthm}
\usepackage{scrextend}
\usepackage{graphicx} 
\usepackage{hyperref}
\usepackage{algorithm}
\usepackage{algpseudocode}
\begin{document}

\title{Multiscale mixing patterns in networks}

\author{Leto Peel}
\email{leto.peel@uclouvain.be}
\affiliation{ICTEAM, Universit\'{e} catholique de Louvain, Louvain-la-Neuve, B-1348, Belgium}
\affiliation{naXys, Université de Namur, Namur, B-5000, Belgium}
\author{Jean-Charles Delvenne} 
\email{jean-charles.delvenne@uclouvain.be}
\affiliation{ICTEAM, Universit\'{e} catholique de Louvain, Louvain-la-Neuve, B-1348, Belgium}
\affiliation{CORE, Universit\'{e} catholique de Louvain, Louvain-la-Neuve, B-1348, Belgium}
\author{Renaud Lambiotte}
\email{renaud.lambiotte@maths.ox.ac.uk}
\affiliation{Mathematical Institute, University of Oxford, Oxford, UK}

\begin{abstract}
Assortative mixing in networks is the tendency for nodes with the same attributes, or metadata, to link to each other.  It is a property often found in social networks manifesting as a higher tendency of links occurring between people with the same age, race, or political belief.  Quantifying the level of assortativity or disassortativity (the preference of linking to nodes with different attributes) can shed light on the organisation of
complex networks.  It is common practice to measure the level of assortativity according to the assortativity coefficient, or modularity in the case of categorical metadata.  This global value is the average level of assortativity across the network and may not be a representative statistic when mixing patterns are heterogeneous.  For example, a social network spanning the globe may exhibit local differences in mixing patterns as a consequence of differences in cultural norms. Here, we introduce an approach to localise this global measure so that we can describe the assortativity, across multiple scales, at the node level.  Consequently we are able to capture and qualitatively evaluate the distribution of mixing patterns in the network. We find that for many real-world networks the distribution of assortativity is skewed, overdispersed and multimodal.  Our  method provides a clearer lens through which we can more closely examine mixing patterns in networks.
\end{abstract}

\maketitle

Networks 
 are used as a common representation for a wide variety of complex systems, spanning social~\cite{krackhardt1999ties, lazega2001collegial, traud2012social}, biological~\cite{brose2005body, jeong2000large} and technological~\cite{albert1999internet, watts1998collective} domains. Nodes are used to represent entities or components of the system and links between them used to indicate pairwise interactions. The link formation processes in these systems are still largely unknown, but the broad variety of observed structures suggest that they are diverse. 
One approach to characterise the network structure is based on the correlation,   
 or \textit{assortative mixing}, of node attributes (or ``metadata'') across edges. This analysis allows us to make generalisations about whether we are more likely to observe links between nodes with the same characteristics (\textit{assortativity}) or between those with different ones (\textit{disassortativity}). 
Social networks frequently contain positive correlations of attribute values across connections~\cite{mcpherson2001birds}.  These correlations occur as a result of the complementary processes of selection (or \textit{``homophily''}) and influence (or \textit{``contagion''})~\cite{aral2009distinguishing}.  
 For example, assortativity has frequently been observed with respect to age, race and social status~\cite{moody2001race}, as well as behavioural patterns such as smoking and drinking habits~\cite{cohen1977sources, kandel1978homophily}.  
Examples of disassortative networks include heterosexual dating networks (gender), ecological food webs (metabolic category), and technological and biological networks (node degree)~\cite{newman2003mixing}. 
It is important to note that just as correlation does not imply causation, 
observations of assortativity are insufficient to imply a specific generative process for the network.

\begin{figure}
\centering
	\includegraphics[width=.9\linewidth]{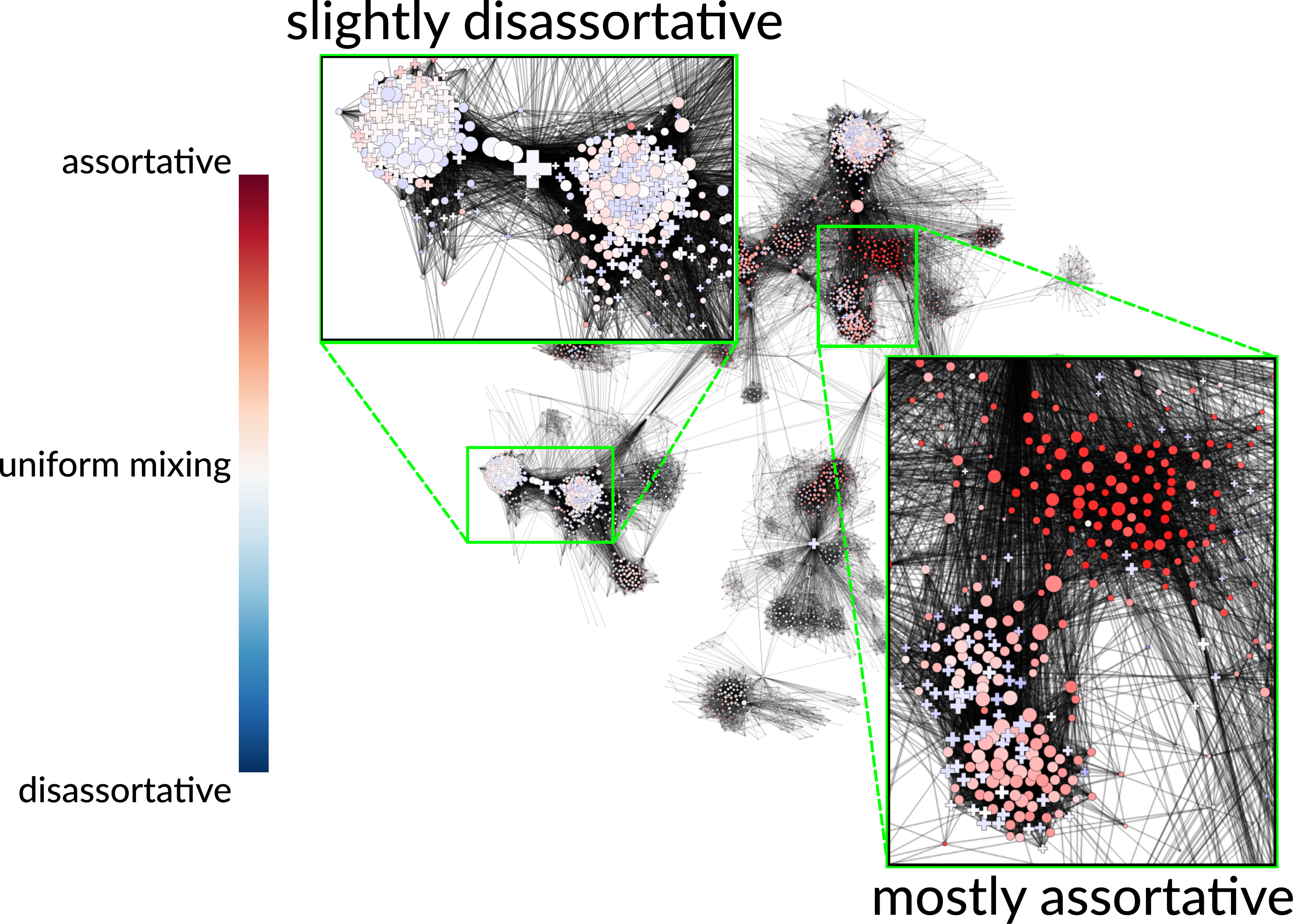}
	\caption{Local assortativity of gender in a sample of Facebook friendships~\cite{mcauley2012learning}. Different regions of the graph exhibit strikingly different patterns, suggesting that a single variable, e.g. global assortativity, would provide a poor description of the system.}
	\label{fig_facebook_ego}
\end{figure}

The standard approach to quantifying the level of assortativity in a network is by calculating the assortativity coefficient~\cite{newman2003mixing}.  Such a summary statistic is useful to capture the average mixing pattern across the whole network.  However, such a generalisation is only really meaningful if it is representative of the population of nodes in the network, i.e., if the assortativity of most individuals is concentrated around the mean. But when networks are heterogeneous and contain diverse mixing patterns, a single global measure may not present an accurate description.
Furthermore it does not provide a means for quantifying the diversity or identifying anomalous or outlier patterns of interaction.

\begin{figure*}
\centering
  \includegraphics[width=.95\linewidth]{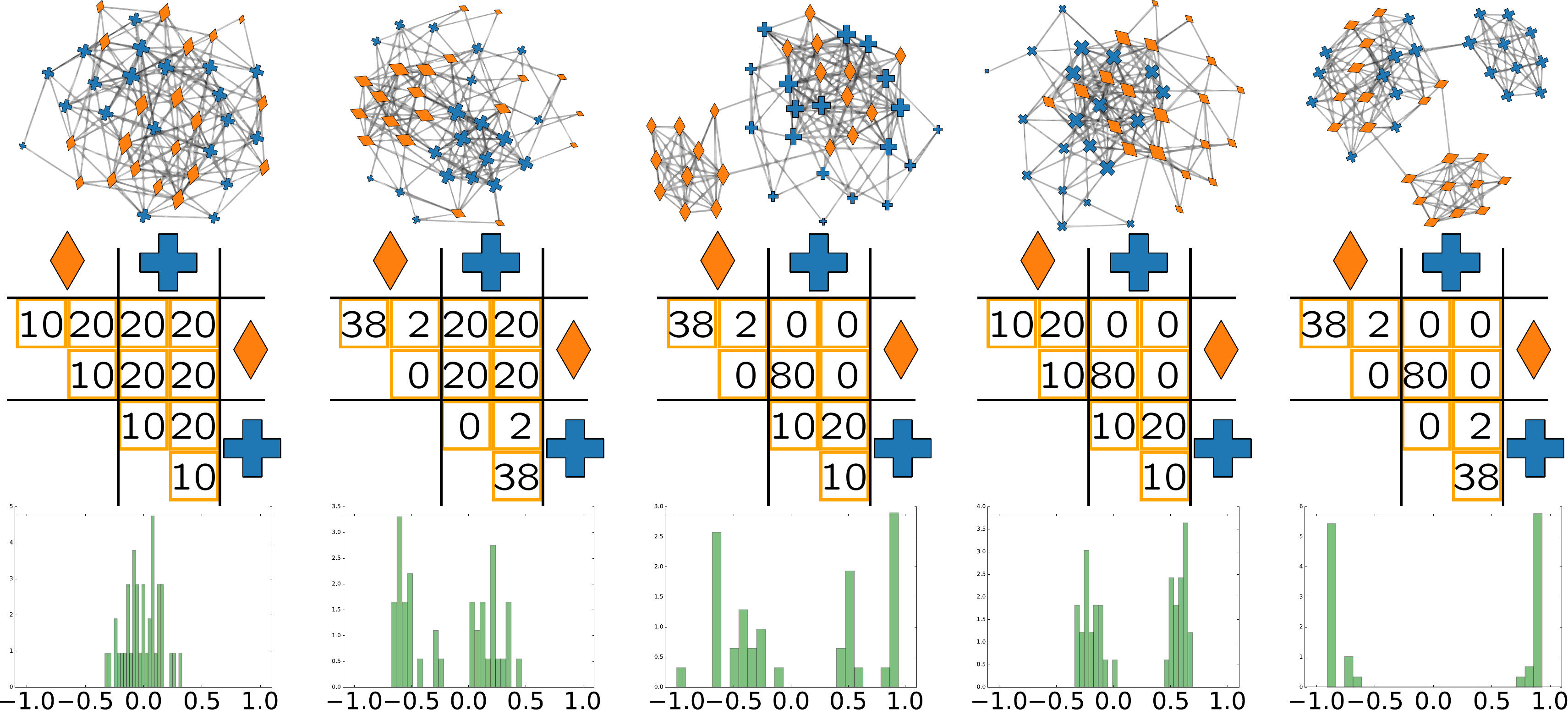}
  \caption{Five networks (top) of $n=40$ nodes and $m=160$ edges with the same global assortativity $r_{\rm{global}}=0$, but with different local mixing patterns as shown by the distributions of $r_{\rm{multi}}$ (bottom). }
  \label{fig_synthNetworks}
\end{figure*}

Quantifying diversity and measuring how mixing may vary across a network becomes a particularly pertinent issue with modern advances in technology that have enabled us to capture, store and process massive-scale networks.  Previously social interaction data was  collected via time-consuming manual processes of conducting surveys or observations. For practical reasons these were often limited to a specific organisation or group~\cite{krackhardt1999ties, lazega2001collegial, sampson1968novitiate, zachary1977information}.  Summarising the pattern of assortative mixing as a single value may be reasonable for these small-scale networks that tend to focus on a single social dimension (e.g., a specific working environment or common interest).  Now, technology such as online social media platforms allow for the automatic 
collection of increasingly larger amounts of social interaction data.  For instance, the largest connected component of the Facebook network was previously reported to account for approximately 10\% of the global population~\cite{ugander2011anatomy}.
These vast multi-dimensional social networks present more opportunities for heterogeneous mixing patterns, which could conceivably arise, for example, due to differences in demographic and cultural backgrounds. 
Figure~\ref{fig_facebook_ego} shows, using the methods we will introduce, an example of this variation in mixing on a subset of nodes in the Facebook social network~\cite{mcauley2012learning}.  A high variation in mixing patterns indicates that the global assortativity may be a poor representation of the entire population. To address this issue, we develop a node-centric measure of the assortativity within a local neighbourhood.
Varying the size of the neighbourhood allows us to interpolate from the mixing pattern between an individual node and its neighbours to the global assortativity coefficient.  In a number of real-world networks we find that the global assortativity is not representative of the collective patterns of mixing.

\section{Mixing in networks}
Currently the standard approach to measure the propensity of links to occur between similar nodes is to use the assortativity coefficient introduced by Newman~\cite{newman2003mixing}.  Here we will focus on undirected networks and categorical node attributes, but assortativity and the methods we propose naturally extend to directed networks and scalar attributes (see Appendix~\ref{app_dir}~\&~\ref{app_sca}). 

The global assortativity coefficient $r_{\textrm{global}}$ for categorical attributes compares the proportion of links connecting nodes  with same attribute value, or \textit{type}, relative to the proportion expected if the edges in the network were randomly rewired. The difference between these proportions is commonly known as modularity $Q$, a measure frequently used in the task of community detection~\cite{newman2004finding}. 
 The assortativity coefficient is  normalised such that $r_{\textrm{global}}=1$ if all edges only connect nodes of the same type (i.e., maximum modularity $Q_{\rm{max}}$) and $r_{\textrm{global}}=0$ if the number of edges is equal to the expected number for a randomly rewired network in which the total number of edges incident on each type of node is held constant. 
 The global assortativity $r_{\textrm{global}}$ is given by~\cite{newman2003mixing}
\begin{equation}
  r_{\textrm{global}} = \frac{Q}{Q_{\rm{max}}} = \frac{\sum_{g}e_{gg}- \sum_{g}a_{g}^{2}}{1-\sum_{g}a_{g}^{2}}  \enspace, \label{eq_assortativity}
\end{equation}
in which $e_{gh}$ is half the proportion of edges in the network that connect nodes with type $y_i=g$ to nodes with type $y_j=h$ (or the proportion of edges if $g=h$) and $a_{g} = \sum_{h} e_{gh} = \sum_{i \in g} k_i / 2m$ is the 
sum of degrees ($k_i$) of nodes with type $g$, normalised by twice the number of edges, $m$.  
We calculate $e_{gh}$ as:
\begin{equation}
    e_{gh} = \frac{1}{2m}\sum_{i: y_i=g}\;\sum_{j: y_j=h} A_{ij}\enspace , \label{eq_egh}     
\end{equation}
where $A_{ij}$ is an element of the adjacency matrix. 
The normalisation constant $Q_{\rm{max}} = 1-\sum_{g}{a_{g}^{2}}$ ensures that the assortativity coefficient lies in the range $ -1 \leq r \leq 1$ (see Appendix~\ref{app_cat}).

\begin{figure*}
\centering
  \includegraphics[width=.85\linewidth]{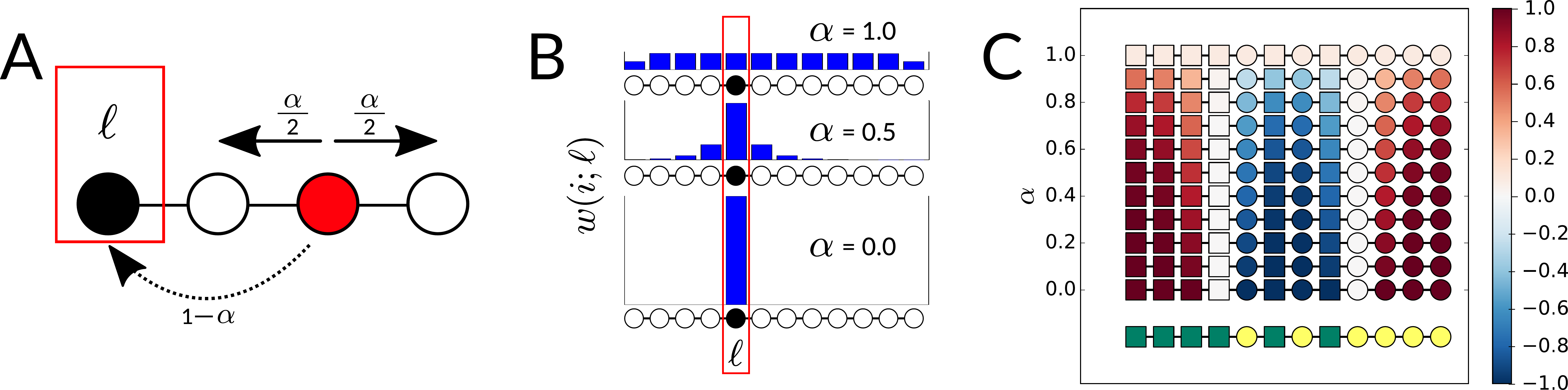}
  \caption{Example of the local assortativity measure for categorical attributes. (\textbf{A}) assortativity is calculated (as in~\eqref{eq_assortativity}) according to the actual proportion of links in the network connecting nodes of the same type relative to the expected proportion of links between nodes of the same type, (\textbf{B}) the nodes in the network are weighted according to a random walk with restart probability of $1-\alpha$, (\textbf{C}) an example of the local assortativity applied to a simple line network with two types of nodes: yellow or green. The blue bars show the stationary distribution ($w(i;\ell)$) of the random walk with restarts at $\ell$ for different values of $\alpha$.  Underneath each distribution the nodes in the line network are coloured according to their local assortativity value.
  }
  \label{fig_ex_local_assort}
\end{figure*}

\subsection{Local patterns of mixing}
The summary statistic $r_{\rm{global}}$ describes the average mixing pattern over the whole network.  But as with all summary statistics there may be cases where it provides a poor representation of the network e.g., if the network contains localised heterogeneous patterns. Figure~\ref{fig_synthNetworks} illustrates an analogy to Anscombe's quartet of bivariate datasets with identical correlation coefficients~\cite{anscombe1973graphs}. Each of the five networks in the top row  have the same number of nodes ($n=40$) and edges ($m=160$) and have been constructed to have the same $r_{\rm{global}}$ with respect to a binary attribute, indicated by a cross ($c$) or a diamond ($d$). All five networks have $m_{cc}+m_{dd}=80$ edges between nodes of the same type and $m_{cd}=80$ edges between nodes of different types, such that each has $r_{\rm{global}}=0$. Local patterns of mixing are formed by splitting each of the types $\{c,d\}$ further into two equally sized subgroups $\{c_1,c_2,d_1,d_2\}$. The middle row depicts the placement of edges within and between the four subgroups.  
Distributing edges uniformly between subgroups creates a network with homogeneous mixing [Fig.~\ref{fig_synthNetworks}(a)].

We propose a local measure of assortativity $r(\ell)$ that captures the mixing pattern within the local neighbourhood of a given node of interest $\ell$.  Trivially one could calculate the local assortativity by adjusting \eqref{eq_assortativity} to only consider the immediate neighbours of $\ell$.  However, taking this approach can encounter problems. For nodes with low degree, we would be calculating assortativity based only on a small sample, providing a potentially poor estimate of the node's mixing preference. And when all of $\ell$'s neighbours are of the same type, then we would assign $r(\ell)=\infty$ because $1-\sum_g a_g^2 = 0$.

We face similar issues in time-series analysis when we wish to interpret how a noisy signal varies over time.  Direct analysis of the series may be more descriptive of the noise process than the underlying signal we are interested in. Averaging over the whole series provides an accurate estimate of the mean, but treats all variation as noise and ignores any important trends.
A common solution to this problem is to use a local filter such as the exponential weighted moving average, in which values further in time from the point of interest are weighted less.
We adopt a similar strategy in calculating the local assortativity.  To make the connection with time-series analysis concrete, we define a random time-series 
where each value is the attribute $y_i$ of a node $i$ visited in a random walk on the graph.
A simple random walker at node $i$ jumps to node $j$ by selecting an outgoing edge with equal probability, $\frac{A_{ij}}{k_{i}}$ and, in an undirected network, the stationary probability  $\pi_{i} = k_{i}/2m$ 
of being at node $i$ is proportional to its degree. Then, every edge of the network is traversed in each direction with equal probability $\pi_i \frac{A_{ij}}{k_{i}} =1/2m$.  In this context, a key observation is that we can equivalently rewrite \eqref{eq_egh} as,
\begin{equation}
	e_{gh}  = \sum_{i: y_i =g} \; \sum_{j: y_j=h} \pi_{i} \frac{A_{ij}}{k_{i}}  \enspace , \label{eq_egh_pi}
\end{equation}
which is the total probability that a simple random walker will jump from a node with type $g$ to one with type $h$. We can then interpret the global assortativity of the network as the autocorrelation (with time lag of 1) of this random time-series (see Appendix~\ref{app_aut} for details).

Global assortativity counts all edges in the network equally just as the stationary random walker visits all edges with equal probability. 
To create our local measure of assortativity we instead reweight the edges in the network based on how local they are to the node of interest,  $\ell$.  We do so by replacing the stationary distribution $\pi$ in \eqref{eq_egh_pi} with an alternative distribution over the nodes $w(i;\ell)$, 
\begin{equation}
  e_{gh}(\ell) = \sum_{i: y_i =g} \; \sum_{j: y_j=h} w(i;\ell) \frac{A_{ij}}{k_{i}}  \enspace,
  \label{eq_egh_w}
\end{equation}
and compare the proportion of links between nodes of the same type in the local neighbourhood to the global value $\nu(\ell) = \sum_{g} (e_{gg}(\ell) - e_{gg})$. Then we can calculate the local assortativity as the deviation from the global assortativity:
\begin{align}
  r(\ell) &= \frac{1}{Q_{\rm{max}}} \left(\nu(\ell) + \sum_{g}e_{gg}- \sum_{g}a_{g}^{2} \right)\\
  &= \frac{1}{Q_{\rm{max}}} \sum_{g} (e_{gg}(\ell) - a_g^2) \enspace .
\end{align}

All that remains is to define a distribution $w(i;\ell)$. We choose the well-known personalised PageRank vector, the stationary distribution $w_{\alpha}(i;\ell)$ of a simple random walk, modified so that at each time step we return to the node of interest $\ell$ with probability $(1-\alpha)$ ~[Fig.~\ref{fig_ex_local_assort}(a)].
In the special case of a network consisting of nodes linked in a line, $w_{\alpha}(i;\ell)$ corresponds to an exponential distribution~[Fig.~\ref{fig_ex_local_assort}(b)] and is analogous to the previously mentioned exponential filter commonly used in time-series analysis. 
 The personalised PageRank vector is an intuitive choice given its role in local community detection~\cite{andersen2006local} and connections to the stochastic block model~\cite{kloumann2016block}. It is however, not the only way to define a local neighbourhood (e.g., a number of graph kernels may be suitable~\cite{fouss2016algorithms}),  but we leave exploration of other neighbourhood functions for future work.

We can now calculate a local assortativity $r_{\alpha}(\ell)$ for each node and use $\alpha$ to interpolate from the trivial local neighbourhood assortativity ($\alpha=0$, the random walker never leaves the initial node) to the global assortativity ($\alpha=1$, the random walker never restarts) $r_{1}(\ell)=r_{\textrm{global}}$~[Fig.~\ref{fig_ex_local_assort}(c)]. We can also view this local assortativity as a (normalised) autocovariance of the random time-series of node attributes, defined as before but now generated by a stationary random walker with restarts and only  when traversing edges of the original network.

\subsection{Choice of $\alpha$}
We can use $\alpha$ to interpolate from the global measure at $\alpha=1$ to the local measure based only on the neighbours of $\ell$ when $\alpha=0$. As previously mentioned, either extreme can be problematic, $r_{1}$ is uniform across network, while $r_{0}$ may be based on a small sample (particularly in the case of low degree nodes) and therefore subject to overfitting.  Moreover, both extremes are blind to the possible existence of coherent regions of assortativity inside the network, as  $r_{1}$ considers the network as a whole while  $r_{0}$ considers the local assortativities of  the nodes as independent entities.

To circumvent these issues, we consider calculating the assortativity across multiple scales by calculating a ``multiscale'' distribution $w_{\rm{multi}}$ by integrating over all possible values of~$\alpha$~\cite{boldi2005totalrank}, 
\begin{equation}
 w_{\rm{multi}}(i;\ell) = \int_0^1 \! w_{\alpha}(i;\ell) \; \rm{d}\alpha \enspace ,
\end{equation}
which is effectively the same as treating $\alpha$ as an unknown with a uniform prior distribution (see Appendix~\ref{app_int} for details).  Using this distribution, we can calculate a multiscale measure $r_{\rm{multi}}$ that captures the assortativity of a given node across all scales.

As a simple demonstration, we return to Figure~\ref{fig_synthNetworks} in which the distribution of $r_{\rm{multi}}$ for each synthetic network is shown in the bottom row.  We see under homogeneous mixing [Fig.~\ref{fig_synthNetworks}(a)] a unimodal distribution peaked around $0$ confirming that the global measure $r_{\rm{global}}$ is representative of the mixing patterns in the network. However, when mixing is heterogeneous [Fig~\ref{fig_synthNetworks}(b--e)] we observe multimodal distributions of $r_{\rm{multi}}$ that allow us to disambiguate between different local mixing patterns.

\begin{figure*}
	\includegraphics[width=\linewidth]{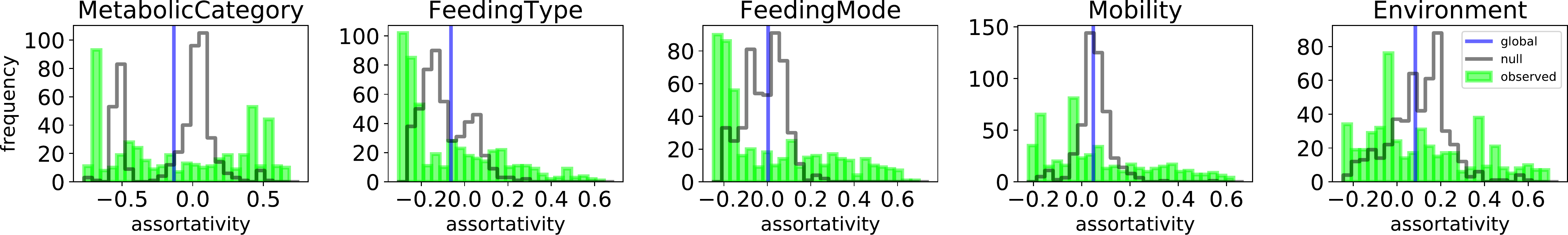}
	\caption{Multiscale assortativity for different attributes in the Weddell Sea Food Web.  The observed distribution is depicted by the solid green bars.  The black outline shows the null distribution. }
	\label{fig_weddell_dist}
\end{figure*}

\section{Real Networks}
Next we use $r_{\rm{multi}}$ to evaluate the mixing patterns in some real networks: an ecological network and set of online social networks. In both cases nodes have multiple attributes assigned to them, providing different dimensions of analysis.

\subsection{Weddell Sea Food Web}
We first examine a network of ecological consumer interactions between species dwelling in the Weddell Sea~\cite{brose2005body}. Figure~\ref{fig_weddell_dist} shows the distributions (\textit{green}) of local assortativity for five different categorical node attributes. For comparison we present a null distribution (\textit{black}) obtained by randomly re-wiring the edges such that the attribute values, degree sequence and global assortativity are all preserved (see Appendix~\ref{app_nul}). In each case we observe skewed and/or multi-modal distributions. The empirical distributions appear overdispersed compared to the null distributions. 

It may be surprising to see that, for some attributes, the null distribution appears to be multi-modal.  Closer inspection reveals that the different modes are correlated with the attribute values and that multi-modality arises from the unbalanced distribution of nodes and incident edges across different node types.  This effect is particularly pronounced for the attribute ``Metabolic Category'' for which we observe two distinct peaks in the distribution.  The larger peak that occurs around $r_{\rm{multi}}\sim 0$ represents all species that belong to the metabolic category \textit{plant} and accounts for the majority ($348/492$) of the species in the network, upon which approximately two thirds 
of the edges are incident. This bias in the distribution of edges across the different node types means that randomly assigned edges are more likely to connect two nodes of the majority class than any other pair of nodes. In fact, it is impossible to assign edges such that nodes in each Metabolic Category exhibit (approximately) the same assortativity as the global value. Specifically, to achieve $r_{\rm{global}}=-0.13$ it is necessary that more than half of the edges 
connect species from different metabolic categories. However, this is impossible for the \textit{plant} category without changing the distribution of edges over categories.

\begin{figure*}[t]
	\includegraphics[width=\linewidth]{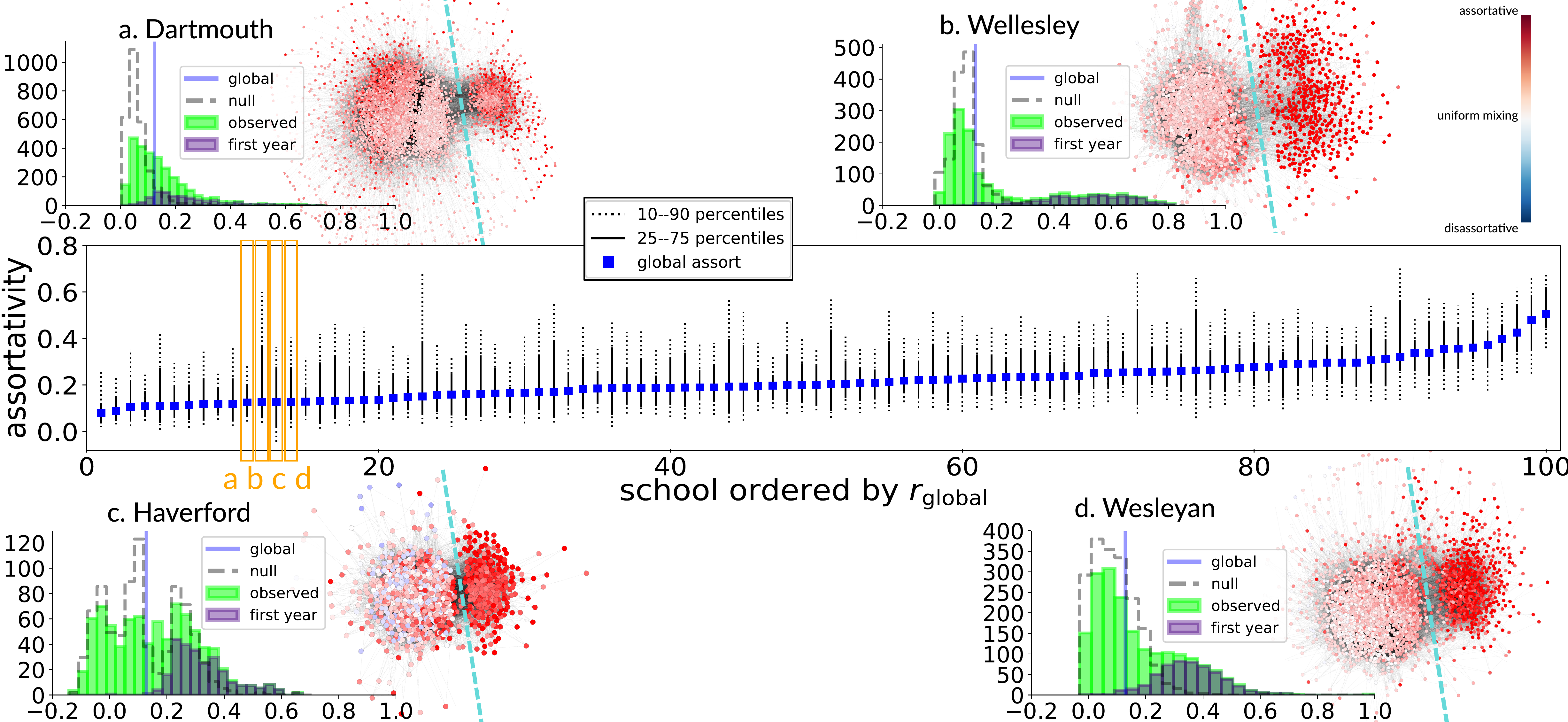}
	\caption{Distributions of the local assortativity by residence (dorm) for each of the schools in the Facebook 100 dataset~\cite{traud2012social}.  Dotted black lines indicate the 10 and 90 percentiles while the solid black lines show the interquartile range.  The global assortativity is indicated by the blue square markers.  The distributions for four schools (Dartmouth, Wesleyan, Wellesley and Haverford) are shown in detail in the surrounding.  Each of them has approximately the same global assortativity ($r_{\rm{global}} \sim 0.13$), but the distributions indicate different levels of heterogeneity in the pattern of mixing by residence. While the distributions are different, there exists a common trend that the first year students tend to be more loosely connected to the rest of the network and exhibit the higher values of assortativity (nodes to the right of the dashed cyan line). }
	\label{fig_fb_assort}
\end{figure*}

\subsection{Facebook 100}
We next consider a set of online social networks collected from the Facebook social media platform at a time when it was only open to 100 US universities~\cite{traud2012social}. The process of incrementally providing these universities access to the platform, meant that at this point in time very few links existed between each of the universities' networks, which provides the opportunities to study each of these social systems in a relatively independent manner. One of the original studies on this dataset examined the assortativity of each demographic attribute in each of the networks~\cite{traud2012social}.  This study found some common patterns that occurred in many of the networks, such as a tendency to be assortative by matriculation year and dormitory of residence, with some variation around the magnitude of assortativity for each of the attributes across the different universities.

\begin{figure*}
	\includegraphics[width=\linewidth]{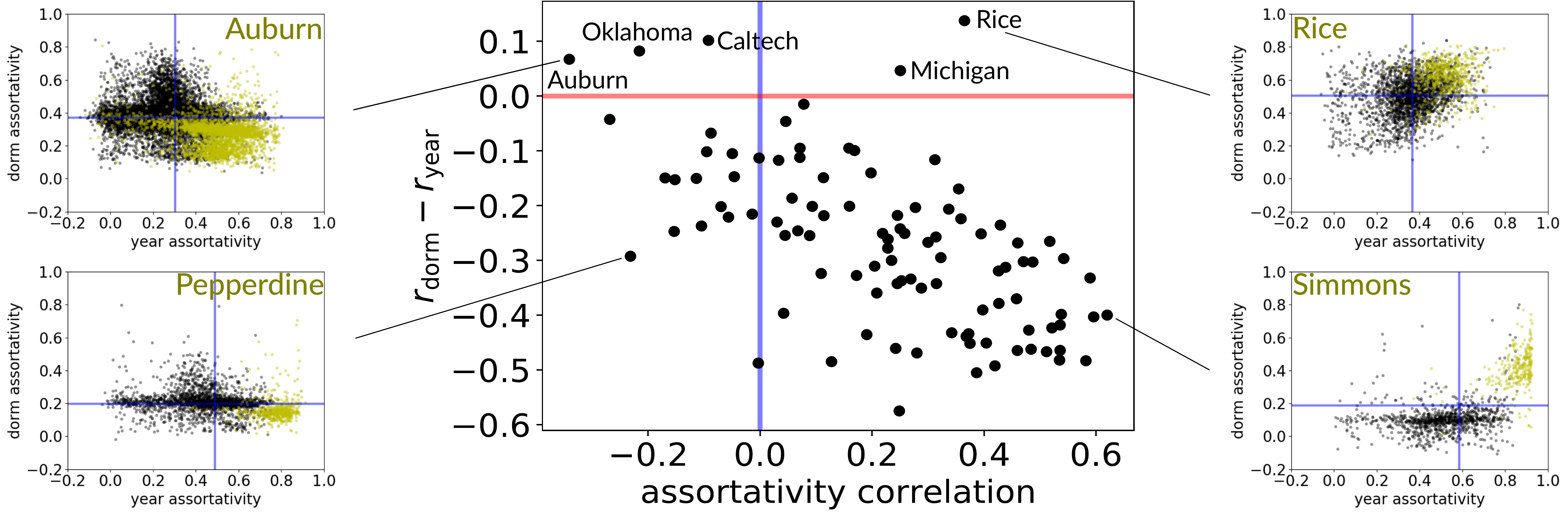}
	\caption{A scatter plot in which each school is a point indicating the correlation of local assortativities by dorm and matriculation year (\textit{x-axis}) and proportion of nodes which are more assortative by dorm than by year (\textit{y-axis}). For reference, the blue vertical line indicates zero assortativity (random mixing) and the red horizontal line indicate zero difference in $r_{\rm{global}}$. Joint distributions of the dorm and year assortativities for four of the schools are shown in the surrounding plots. Blue lines indicate $r_{\rm{global}}$. }
	\label{fig_fb_corr}
\end{figure*}

In this case it makes sense to analyse the universities separately since it is reasonable to assume that university membership played an important and restricting role in the organisation of the network.  However, a modern version of this dataset might contain a higher density of inter-university links, making it less reasonable to treat them independently; in general partitioning networks based on attributes without careful considerations can be problematic~\cite{peel2017ground}.

Figure~\ref{fig_fb_assort} depicts the distributions of $r_{\rm{multi}}$ for each of the 100 networks according to dormitory.  For many of the networks we observe a positively-skewed distribution.  The surrounding subplots show details for four universities with approximately the same global assortativity $r_{\rm{global}} \sim 0.13$, but with qualitatively different distributions of $r_{\rm{multi}}$.  Common across these distributions is that  all of the empirical distributions exhibit a positive skew beyond that of the null distribution.  Closer inspection reveals that in all four networks the nodes associated with a higher local assortativity belong to a community of nodes more loosely connected to the rest of the network.  These nodes also correspond to  first-year students, which suggests that residence is more relevant to friendship among new students than it is for the rest of the student body. 
We see this pattern in many of the other schools too, first-year students are more assortative by year (for all schools except one) and by residence (more than 75\% of schools), see~Figs.~\ref{fig_fb_year}~\&~\ref{fig_fb_dorm} in the Appendix for details.

We can also use local assortativity to compare how the mixing of multiple attributes covary across a network. This may be of interest as a positive correlation could suggest a relationship between attributes, while a negative correlation indicates that assortativity of one attribute may replace the assortativity of another. 
Note that differences in normalisation between attributes mean that the actual values may not be directly comparable, which is why we focus on correlation.
Figure~\ref{fig_fb_corr} compares $r_{\rm{multi}}$ for year of study and place of residence.  The central scatter plot shows, for each university: the correlation between local assortativities of the two attributes (x-axis) against the difference in the two global assortativities for each network (y-axis), which was previously the only way to compare assortativities~\cite{traud2012social}.
The four surrounding sub-plots show the joint distribution of year and dorm local assortativity for specific universities. The yellow points indicate students in their first year.  In most universities we observed that first-year students were the most assortative by either year, residence or both.  In both Auburn and Pepperdine there is a negative correlation between year and dorm assortativity suggesting that many friendships are associated with either being in the same year \textit{or} from sharing a dorm.

For Simmons and Rice we observe a positive correlation between dorm and year local assortativity. However, in Simmons we see that the first-year students form a separate cluster, while in Rice they are much more interspersed.  This difference may relate to how students are placed in university dorms.  At Simmons all first year students live on campus\footnote{source:\url{http://www.simmons.edu/student-life/life-at-simmons/housing/residence-halls}} and form the majority residents in the few dorms they occupy.  Rice houses their new intakes according to a different strategy, by placing them evenly spread across all the available dorms. The fact that students are mixed across years and that the vast majority (almost 78\%\footnote{source:\url{http://campushousing.rice.edu/}}) of students reside in university accommodation, offers a possible explanation for why we observe a smooth variation in values of assortativity without a distinction between new students and the rest of the population.

\section{Discussion}
Characterising the level of assortativity plays an important role in understanding the organisation of complex systems.  
However, the global assortativity may not be representative given the variation present in the network. We have shown that the distribution of mixing in real networks can be skewed, overdispersed and possibly multimodal.  In fact, for certain network configurations we have seen that a unimodal distribution may not even be possible.

As network data grow bigger there is a greater possibility for heterogeneous sub-groups to co-exist within the overall population. The presence of these sub-populations adds further to the ongoing discussions of the interplay between node metadata and network structure~\cite{peel2017ground} and suggests that while we may observe a relationship between particular node properties and existence of links in part of a network, it does not imply that this relationship exists across the network as a whole.  
This heterogeneity has implications on how we make generalisations in network data, as what we observe in a subgraph might not necessarily apply to the rest of the network. However, it may also present new opportunities too.  Recent results show that with an appropriately constructed learning algorithm it is still possible to make accurate predictions about node attributes in networks with heterogeneous mixing patterns~\cite{peel2017graph} and in some cases even utilise the heterogeneity to further improve performance~\cite{altenburger2017bias}. Quantifying local assortativity offers a new dimension to study this predictive performance.
Heterogeneous mixing also offers a potential new perspective for the community detection problem~\cite{schaub2017many}, i.e., to identify sets of nodes with similar assortativity, which may be useful in the study of ``echo chambers'' in social networks~\cite{colleoni2014echo}.

Our approach to quantifying local mixing could easily be applied to any global network measure,  such as clustering coefficient or mean degree. It may also be used to capture the local correlation between node attributes and their degree, a relationship that plays a definitive role in network phenomena such as the majority illusion~\cite{lerman2016majority} and the generalised friendship paradox~\cite{eom2014generalized}.

\section*{Acknowledgements}
This work was supported by ARC (Federation Wallonia-Brussels) [LP, JCD, RL] and FNRS-F.R.S. [LP].

\bibliography{refs}

\begin{thebibliography}{38}%
\makeatletter
\providecommand \@ifxundefined [1]{%
 \@ifx{#1\undefined}
}%
\providecommand \@ifnum [1]{%
 \ifnum #1\expandafter \@firstoftwo
 \else \expandafter \@secondoftwo
 \fi
}%
\providecommand \@ifx [1]{%
 \ifx #1\expandafter \@firstoftwo
 \else \expandafter \@secondoftwo
 \fi
}%
\providecommand \natexlab [1]{#1}%
\providecommand \enquote  [1]{``#1''}%
\providecommand \bibnamefont  [1]{#1}%
\providecommand \bibfnamefont [1]{#1}%
\providecommand \citenamefont [1]{#1}%
\providecommand \href@noop [0]{\@secondoftwo}%
\providecommand \href [0]{\begingroup \@sanitize@url \@href}%
\providecommand \@href[1]{\@@startlink{#1}\@@href}%
\providecommand \@@href[1]{\endgroup#1\@@endlink}%
\providecommand \@sanitize@url [0]{\catcode `\\12\catcode `\$12\catcode
  `\&12\catcode `\#12\catcode `\^12\catcode `\_12\catcode `\%12\relax}%
\providecommand \@@startlink[1]{}%
\providecommand \@@endlink[0]{}%
\providecommand \url  [0]{\begingroup\@sanitize@url \@url }%
\providecommand \@url [1]{\endgroup\@href {#1}{\urlprefix }}%
\providecommand \urlprefix  [0]{URL }%
\providecommand \Eprint [0]{\href }%
\providecommand \doibase [0]{http://dx.doi.org/}%
\providecommand \selectlanguage [0]{\@gobble}%
\providecommand \bibinfo  [0]{\@secondoftwo}%
\providecommand \bibfield  [0]{\@secondoftwo}%
\providecommand \translation [1]{[#1]}%
\providecommand \BibitemOpen [0]{}%
\providecommand \bibitemStop [0]{}%
\providecommand \bibitemNoStop [0]{.\EOS\space}%
\providecommand \EOS [0]{\spacefactor3000\relax}%
\providecommand \BibitemShut  [1]{\csname bibitem#1\endcsname}%
\let\auto@bib@innerbib\@empty
\bibitem [{\citenamefont {Krackhardt}(1999)}]{krackhardt1999ties}%
  \BibitemOpen
  \bibfield  {author} {\bibinfo {author} {\bibfnamefont {David}\ \bibnamefont
  {Krackhardt}},\ }\bibfield  {title} {\enquote {\bibinfo {title} {The ties
  that torture: Simmelian tie analysis in organizations},}\ }\href@noop {}
  {\bibfield  {journal} {\bibinfo  {journal} {Research in the Sociology of
  Organizations}\ }\textbf {\bibinfo {volume} {16}},\ \bibinfo {pages}
  {183--210} (\bibinfo {year} {1999})}\BibitemShut {NoStop}%
\bibitem [{\citenamefont {Lazega}(2001)}]{lazega2001collegial}%
  \BibitemOpen
  \bibfield  {author} {\bibinfo {author} {\bibfnamefont {Emmanuel}\
  \bibnamefont {Lazega}},\ }\href@noop {} {\emph {\bibinfo {title} {The
  collegial phenomenon: The social mechanisms of cooperation among peers in a
  corporate law partnership}}}\ (\bibinfo  {publisher} {Oxford University
  Press},\ \bibinfo {year} {2001})\BibitemShut {NoStop}%
\bibitem [{\citenamefont {Traud}\ \emph {et~al.}(2012)\citenamefont {Traud},
  \citenamefont {Mucha},\ and\ \citenamefont {Porter}}]{traud2012social}%
  \BibitemOpen
  \bibfield  {author} {\bibinfo {author} {\bibfnamefont {Amanda~L}\
  \bibnamefont {Traud}}, \bibinfo {author} {\bibfnamefont {Peter~J}\
  \bibnamefont {Mucha}}, \ and\ \bibinfo {author} {\bibfnamefont {Mason~A}\
  \bibnamefont {Porter}},\ }\bibfield  {title} {\enquote {\bibinfo {title}
  {Social structure of facebook networks},}\ }\href@noop {} {\bibfield
  {journal} {\bibinfo  {journal} {Physica A}\ }\textbf {\bibinfo {volume}
  {391}},\ \bibinfo {pages} {4165--4180} (\bibinfo {year} {2012})}\BibitemShut
  {NoStop}%
\bibitem [{\citenamefont {Brose}\ \emph {et~al.}(2005)\citenamefont {Brose}
  \emph {et~al.}}]{brose2005body}%
  \BibitemOpen
  \bibfield  {author} {\bibinfo {author} {\bibfnamefont {Ulrich}\ \bibnamefont
  {Brose}} \emph {et~al.},\ }\bibfield  {title} {\enquote {\bibinfo {title}
  {{Body sizes of consumers and their resources}},}\ }\href@noop {} {\bibfield
  {journal} {\bibinfo  {journal} {Ecology}\ }\textbf {\bibinfo {volume} {86}},\
  \bibinfo {pages} {2545} (\bibinfo {year} {2005})}\BibitemShut {NoStop}%
\bibitem [{\citenamefont {Jeong}\ \emph {et~al.}(2000)\citenamefont {Jeong},
  \citenamefont {Tombor}, \citenamefont {Albert}, \citenamefont {Oltvai},\ and\
  \citenamefont {Barabasi}}]{jeong2000large}%
  \BibitemOpen
  \bibfield  {author} {\bibinfo {author} {\bibfnamefont {H}~\bibnamefont
  {Jeong}}, \bibinfo {author} {\bibfnamefont {B}~\bibnamefont {Tombor}},
  \bibinfo {author} {\bibfnamefont {R}~\bibnamefont {Albert}}, \bibinfo
  {author} {\bibfnamefont {ZN}~\bibnamefont {Oltvai}}, \ and\ \bibinfo {author}
  {\bibfnamefont {AL}~\bibnamefont {Barabasi}},\ }\bibfield  {title} {\enquote
  {\bibinfo {title} {The large-scale organization of metabolic networks},}\
  }\href@noop {} {\bibfield  {journal} {\bibinfo  {journal} {Nature}\ }\textbf
  {\bibinfo {volume} {407}},\ \bibinfo {pages} {651} (\bibinfo {year}
  {2000})}\BibitemShut {NoStop}%
\bibitem [{\citenamefont {Albert}\ \emph {et~al.}(1999)\citenamefont {Albert},
  \citenamefont {Jeong},\ and\ \citenamefont
  {Barab{\'a}si}}]{albert1999internet}%
  \BibitemOpen
  \bibfield  {author} {\bibinfo {author} {\bibfnamefont {R{\'e}ka}\
  \bibnamefont {Albert}}, \bibinfo {author} {\bibfnamefont {Hawoong}\
  \bibnamefont {Jeong}}, \ and\ \bibinfo {author} {\bibfnamefont
  {Albert-L{\'a}szl{\'o}}\ \bibnamefont {Barab{\'a}si}},\ }\bibfield  {title}
  {\enquote {\bibinfo {title} {Internet: Diameter of the world-wide web},}\
  }\href@noop {} {\bibfield  {journal} {\bibinfo  {journal} {Nature}\ }\textbf
  {\bibinfo {volume} {401}},\ \bibinfo {pages} {130} (\bibinfo {year}
  {1999})}\BibitemShut {NoStop}%
\bibitem [{\citenamefont {Watts}\ and\ \citenamefont
  {Strogatz}(1998)}]{watts1998collective}%
  \BibitemOpen
  \bibfield  {author} {\bibinfo {author} {\bibfnamefont {Duncan~J}\
  \bibnamefont {Watts}}\ and\ \bibinfo {author} {\bibfnamefont {Steven~H}\
  \bibnamefont {Strogatz}},\ }\bibfield  {title} {\enquote {\bibinfo {title}
  {Collective dynamics of 'small-world' networks},}\ }\href@noop {} {\bibfield
  {journal} {\bibinfo  {journal} {Nature}\ }\textbf {\bibinfo {volume} {393}},\
  \bibinfo {pages} {440} (\bibinfo {year} {1998})}\BibitemShut {NoStop}%
\bibitem [{\citenamefont {McPherson}\ \emph {et~al.}(2001)\citenamefont
  {McPherson}, \citenamefont {Smith-Lovin},\ and\ \citenamefont
  {Cook}}]{mcpherson2001birds}%
  \BibitemOpen
  \bibfield  {author} {\bibinfo {author} {\bibfnamefont {Miller}\ \bibnamefont
  {McPherson}}, \bibinfo {author} {\bibfnamefont {Lynn}\ \bibnamefont
  {Smith-Lovin}}, \ and\ \bibinfo {author} {\bibfnamefont {James~M}\
  \bibnamefont {Cook}},\ }\bibfield  {title} {\enquote {\bibinfo {title} {Birds
  of a feather: Homophily in social networks},}\ }\href@noop {} {\bibfield
  {journal} {\bibinfo  {journal} {Annu. Rev. Sociol.}\ }\textbf {\bibinfo
  {volume} {27}},\ \bibinfo {pages} {415--444} (\bibinfo {year}
  {2001})}\BibitemShut {NoStop}%
\bibitem [{\citenamefont {Aral}\ \emph {et~al.}(2009)\citenamefont {Aral},
  \citenamefont {Muchnik},\ and\ \citenamefont
  {Sundararajan}}]{aral2009distinguishing}%
  \BibitemOpen
  \bibfield  {author} {\bibinfo {author} {\bibfnamefont {Sinan}\ \bibnamefont
  {Aral}}, \bibinfo {author} {\bibfnamefont {Lev}\ \bibnamefont {Muchnik}}, \
  and\ \bibinfo {author} {\bibfnamefont {Arun}\ \bibnamefont {Sundararajan}},\
  }\bibfield  {title} {\enquote {\bibinfo {title} {Distinguishing
  influence-based contagion from homophily-driven diffusion in dynamic
  networks},}\ }\href@noop {} {\bibfield  {journal} {\bibinfo  {journal} {Proc.
  Natl. Acad. Sci.}\ }\textbf {\bibinfo {volume} {106}},\ \bibinfo {pages}
  {21544--21549} (\bibinfo {year} {2009})}\BibitemShut {NoStop}%
\bibitem [{\citenamefont {Moody}(2001)}]{moody2001race}%
  \BibitemOpen
  \bibfield  {author} {\bibinfo {author} {\bibfnamefont {James}\ \bibnamefont
  {Moody}},\ }\bibfield  {title} {\enquote {\bibinfo {title} {Race, school
  integration, and friendship segregation in america},}\ }\href@noop {}
  {\bibfield  {journal} {\bibinfo  {journal} {Am. J. Sociol.}\ }\textbf
  {\bibinfo {volume} {107}},\ \bibinfo {pages} {679--716} (\bibinfo {year}
  {2001})}\BibitemShut {NoStop}%
\bibitem [{\citenamefont {Cohen}(1977)}]{cohen1977sources}%
  \BibitemOpen
  \bibfield  {author} {\bibinfo {author} {\bibfnamefont {Jere~M}\ \bibnamefont
  {Cohen}},\ }\bibfield  {title} {\enquote {\bibinfo {title} {Sources of peer
  group homogeneity},}\ }\href@noop {} {\bibfield  {journal} {\bibinfo
  {journal} {Sociol Educ}\ ,\ \bibinfo {pages} {227--241}} (\bibinfo {year}
  {1977})}\BibitemShut {NoStop}%
\bibitem [{\citenamefont {Kandel}(1978)}]{kandel1978homophily}%
  \BibitemOpen
  \bibfield  {author} {\bibinfo {author} {\bibfnamefont {Denise~B}\
  \bibnamefont {Kandel}},\ }\bibfield  {title} {\enquote {\bibinfo {title}
  {Homophily, selection, and socialization in adolescent friendships},}\
  }\href@noop {} {\bibfield  {journal} {\bibinfo  {journal} {Am. J. Sociol.}\
  }\textbf {\bibinfo {volume} {84}},\ \bibinfo {pages} {427--436} (\bibinfo
  {year} {1978})}\BibitemShut {NoStop}%
\bibitem [{\citenamefont {Newman}(2003)}]{newman2003mixing}%
  \BibitemOpen
  \bibfield  {author} {\bibinfo {author} {\bibfnamefont {Mark~EJ}\ \bibnamefont
  {Newman}},\ }\bibfield  {title} {\enquote {\bibinfo {title} {Mixing patterns
  in networks},}\ }\href@noop {} {\bibfield  {journal} {\bibinfo  {journal}
  {Phys. Rev. E}\ }\textbf {\bibinfo {volume} {67}},\ \bibinfo {pages} {026126}
  (\bibinfo {year} {2003})}\BibitemShut {NoStop}%
\bibitem [{\citenamefont {McAuley}\ and\ \citenamefont
  {Leskovec}(2012)}]{mcauley2012learning}%
  \BibitemOpen
  \bibfield  {author} {\bibinfo {author} {\bibfnamefont {Julian~J}\
  \bibnamefont {McAuley}}\ and\ \bibinfo {author} {\bibfnamefont {Jure}\
  \bibnamefont {Leskovec}},\ }\bibfield  {title} {\enquote {\bibinfo {title}
  {Learning to discover social circles in ego networks},}\ }in\ \href@noop {}
  {\emph {\bibinfo {booktitle} {Adv. Neur. In. 25}}},\ Vol.\ \bibinfo {volume}
  {2012}\ (\bibinfo {year} {2012})\ pp.\ \bibinfo {pages}
  {539--547}\BibitemShut {NoStop}%
\bibitem [{\citenamefont {Sampson}(1968)}]{sampson1968novitiate}%
  \BibitemOpen
  \bibfield  {author} {\bibinfo {author} {\bibfnamefont {Samuel~F}\
  \bibnamefont {Sampson}},\ }\emph {\bibinfo {title} {A novitiate in a period
  of change: An experimental and case study of social relationships}},\
  \href@noop {} {Ph.D. thesis},\ \bibinfo  {school} {Cornell University Ithaca,
  NY} (\bibinfo {year} {1968})\BibitemShut {NoStop}%
\bibitem [{\citenamefont {Zachary}(1977)}]{zachary1977information}%
  \BibitemOpen
  \bibfield  {author} {\bibinfo {author} {\bibfnamefont {Wayne~W}\ \bibnamefont
  {Zachary}},\ }\bibfield  {title} {\enquote {\bibinfo {title} {An information
  flow model for conflict and fission in small groups},}\ }\href@noop {}
  {\bibfield  {journal} {\bibinfo  {journal} {J. Anthropol. Res.}\ ,\ \bibinfo
  {pages} {452--473}} (\bibinfo {year} {1977})}\BibitemShut {NoStop}%
\bibitem [{\citenamefont {Ugander}\ \emph {et~al.}(2011)\citenamefont
  {Ugander}, \citenamefont {Karrer}, \citenamefont {Backstrom},\ and\
  \citenamefont {Marlow}}]{ugander2011anatomy}%
  \BibitemOpen
  \bibfield  {author} {\bibinfo {author} {\bibfnamefont {Johan}\ \bibnamefont
  {Ugander}}, \bibinfo {author} {\bibfnamefont {Brian}\ \bibnamefont {Karrer}},
  \bibinfo {author} {\bibfnamefont {Lars}\ \bibnamefont {Backstrom}}, \ and\
  \bibinfo {author} {\bibfnamefont {Cameron}\ \bibnamefont {Marlow}},\
  }\bibfield  {title} {\enquote {\bibinfo {title} {The anatomy of the facebook
  social graph},}\ }\href@noop {} {\bibfield  {journal} {\bibinfo  {journal}
  {preprint arXiv:1111.4503}\ } (\bibinfo {year} {2011})}\BibitemShut {NoStop}%
\bibitem [{\citenamefont {Newman}\ and\ \citenamefont
  {Girvan}(2004)}]{newman2004finding}%
  \BibitemOpen
  \bibfield  {author} {\bibinfo {author} {\bibfnamefont {Mark~EJ}\ \bibnamefont
  {Newman}}\ and\ \bibinfo {author} {\bibfnamefont {Michelle}\ \bibnamefont
  {Girvan}},\ }\bibfield  {title} {\enquote {\bibinfo {title} {Finding and
  evaluating community structure in networks},}\ }\href@noop {} {\bibfield
  {journal} {\bibinfo  {journal} {Phys. rev. E}\ }\textbf {\bibinfo {volume}
  {69}},\ \bibinfo {pages} {026113} (\bibinfo {year} {2004})}\BibitemShut
  {NoStop}%
\bibitem [{\citenamefont {Anscombe}(1973)}]{anscombe1973graphs}%
  \BibitemOpen
  \bibfield  {author} {\bibinfo {author} {\bibfnamefont {Francis~J}\
  \bibnamefont {Anscombe}},\ }\bibfield  {title} {\enquote {\bibinfo {title}
  {Graphs in statistical analysis},}\ }\href@noop {} {\bibfield  {journal}
  {\bibinfo  {journal} {Am. Stat.}\ }\textbf {\bibinfo {volume} {27}},\
  \bibinfo {pages} {17--21} (\bibinfo {year} {1973})}\BibitemShut {NoStop}%
\bibitem [{\citenamefont {Andersen}\ \emph {et~al.}(2006)\citenamefont
  {Andersen}, \citenamefont {Chung},\ and\ \citenamefont
  {Lang}}]{andersen2006local}%
  \BibitemOpen
  \bibfield  {author} {\bibinfo {author} {\bibfnamefont {Reid}\ \bibnamefont
  {Andersen}}, \bibinfo {author} {\bibfnamefont {Fan}\ \bibnamefont {Chung}}, \
  and\ \bibinfo {author} {\bibfnamefont {Kevin}\ \bibnamefont {Lang}},\
  }\bibfield  {title} {\enquote {\bibinfo {title} {Local graph partitioning
  using pagerank vectors},}\ }in\ \href@noop {} {\emph {\bibinfo {booktitle}
  {Foundations of Computer Science (FOCS'06)}}}\ (\bibinfo {organization}
  {IEEE},\ \bibinfo {year} {2006})\ pp.\ \bibinfo {pages}
  {475--486}\BibitemShut {NoStop}%
\bibitem [{\citenamefont {Kloumann}\ \emph {et~al.}(2016)\citenamefont
  {Kloumann}, \citenamefont {Ugander},\ and\ \citenamefont
  {Kleinberg}}]{kloumann2016block}%
  \BibitemOpen
  \bibfield  {author} {\bibinfo {author} {\bibfnamefont {Isabel~M}\
  \bibnamefont {Kloumann}}, \bibinfo {author} {\bibfnamefont {Johan}\
  \bibnamefont {Ugander}}, \ and\ \bibinfo {author} {\bibfnamefont {Jon}\
  \bibnamefont {Kleinberg}},\ }\bibfield  {title} {\enquote {\bibinfo {title}
  {Block models and personalized pagerank},}\ }\href@noop {} {\bibfield
  {journal} {\bibinfo  {journal} {Proc. Natl. Acad. Sci.}\ }\textbf {\bibinfo
  {volume} {114}},\ \bibinfo {pages} {33--38} (\bibinfo {year}
  {2016})}\BibitemShut {NoStop}%
\bibitem [{\citenamefont {Fouss}\ \emph {et~al.}(2016)\citenamefont {Fouss},
  \citenamefont {Saerens},\ and\ \citenamefont {Shimbo}}]{fouss2016algorithms}%
  \BibitemOpen
  \bibfield  {author} {\bibinfo {author} {\bibfnamefont {Fran{\c{c}}ois}\
  \bibnamefont {Fouss}}, \bibinfo {author} {\bibfnamefont {Marco}\ \bibnamefont
  {Saerens}}, \ and\ \bibinfo {author} {\bibfnamefont {Masashi}\ \bibnamefont
  {Shimbo}},\ }\href@noop {} {\emph {\bibinfo {title} {Algorithms and models
  for network data and link analysis}}}\ (\bibinfo  {publisher} {Cambridge
  University Press},\ \bibinfo {year} {2016})\BibitemShut {NoStop}%
\bibitem [{\citenamefont {Boldi}(2005)}]{boldi2005totalrank}%
  \BibitemOpen
  \bibfield  {author} {\bibinfo {author} {\bibfnamefont {Paolo}\ \bibnamefont
  {Boldi}},\ }\bibfield  {title} {\enquote {\bibinfo {title} {Totalrank:
  Ranking without damping},}\ }in\ \href@noop {} {\emph {\bibinfo {booktitle}
  {Special interest tracks and posters of the 14th international conference on
  World Wide Web (WWW)}}}\ (\bibinfo {organization} {ACM},\ \bibinfo {year}
  {2005})\ pp.\ \bibinfo {pages} {898--899}\BibitemShut {NoStop}%
\bibitem [{\citenamefont {Peel}\ \emph {et~al.}(2017)\citenamefont {Peel},
  \citenamefont {Larremore},\ and\ \citenamefont {Clauset}}]{peel2017ground}%
  \BibitemOpen
  \bibfield  {author} {\bibinfo {author} {\bibfnamefont {Leto}\ \bibnamefont
  {Peel}}, \bibinfo {author} {\bibfnamefont {Daniel~B}\ \bibnamefont
  {Larremore}}, \ and\ \bibinfo {author} {\bibfnamefont {Aaron}\ \bibnamefont
  {Clauset}},\ }\bibfield  {title} {\enquote {\bibinfo {title} {The ground
  truth about metadata and community detection in networks},}\ }\href@noop {}
  {\bibfield  {journal} {\bibinfo  {journal} {Sci. Adv.}\ }\textbf {\bibinfo
  {volume} {3}},\ \bibinfo {pages} {e1602548} (\bibinfo {year}
  {2017})}\BibitemShut {NoStop}%
\bibitem [{\citenamefont {Peel}(2017)}]{peel2017graph}%
  \BibitemOpen
  \bibfield  {author} {\bibinfo {author} {\bibfnamefont {Leto}\ \bibnamefont
  {Peel}},\ }\bibfield  {title} {\enquote {\bibinfo {title} {Graph-based
  semi-supervised learning for relational networks},}\ }in\ \href@noop {}
  {\emph {\bibinfo {booktitle} {SIAM International Conference on Data
  Mining}}}\ (\bibinfo {organization} {SIAM},\ \bibinfo {year} {2017})\ pp.\
  \bibinfo {pages} {435--443}\BibitemShut {NoStop}%
\bibitem [{\citenamefont {Altenburger}\ and\ \citenamefont
  {Ugander}(2017)}]{altenburger2017bias}%
  \BibitemOpen
  \bibfield  {author} {\bibinfo {author} {\bibfnamefont {Kristen~M}\
  \bibnamefont {Altenburger}}\ and\ \bibinfo {author} {\bibfnamefont {Johan}\
  \bibnamefont {Ugander}},\ }\bibfield  {title} {\enquote {\bibinfo {title}
  {Bias and variance in the social structure of gender},}\ }\href@noop {}
  {\bibfield  {journal} {\bibinfo  {journal} {preprint arXiv:1705.04774}\ }
  (\bibinfo {year} {2017})}\BibitemShut {NoStop}%
\bibitem [{\citenamefont {Schaub}\ \emph {et~al.}(2017)\citenamefont {Schaub},
  \citenamefont {Delvenne}, \citenamefont {Rosvall},\ and\ \citenamefont
  {Lambiotte}}]{schaub2017many}%
  \BibitemOpen
  \bibfield  {author} {\bibinfo {author} {\bibfnamefont {Michael~T}\
  \bibnamefont {Schaub}}, \bibinfo {author} {\bibfnamefont {Jean-Charles}\
  \bibnamefont {Delvenne}}, \bibinfo {author} {\bibfnamefont {Martin}\
  \bibnamefont {Rosvall}}, \ and\ \bibinfo {author} {\bibfnamefont {Renaud}\
  \bibnamefont {Lambiotte}},\ }\bibfield  {title} {\enquote {\bibinfo {title}
  {The many facets of community detection in complex networks},}\ }\href@noop
  {} {\bibfield  {journal} {\bibinfo  {journal} {Applied Network Science}\
  }\textbf {\bibinfo {volume} {2}},\ \bibinfo {pages} {4} (\bibinfo {year}
  {2017})}\BibitemShut {NoStop}%
\bibitem [{\citenamefont {Colleoni}\ \emph {et~al.}(2014)\citenamefont
  {Colleoni}, \citenamefont {Rozza},\ and\ \citenamefont
  {Arvidsson}}]{colleoni2014echo}%
  \BibitemOpen
  \bibfield  {author} {\bibinfo {author} {\bibfnamefont {Elanor}\ \bibnamefont
  {Colleoni}}, \bibinfo {author} {\bibfnamefont {Alessandro}\ \bibnamefont
  {Rozza}}, \ and\ \bibinfo {author} {\bibfnamefont {Adam}\ \bibnamefont
  {Arvidsson}},\ }\bibfield  {title} {\enquote {\bibinfo {title} {Echo chamber
  or public sphere? predicting political orientation and measuring political
  homophily in twitter using big data},}\ }\href@noop {} {\bibfield  {journal}
  {\bibinfo  {journal} {J. Commun.}\ }\textbf {\bibinfo {volume} {64}},\
  \bibinfo {pages} {317--332} (\bibinfo {year} {2014})}\BibitemShut {NoStop}%
\bibitem [{\citenamefont {Lerman}\ \emph {et~al.}(2016)\citenamefont {Lerman},
  \citenamefont {Yan},\ and\ \citenamefont {Wu}}]{lerman2016majority}%
  \BibitemOpen
  \bibfield  {author} {\bibinfo {author} {\bibfnamefont {Kristina}\
  \bibnamefont {Lerman}}, \bibinfo {author} {\bibfnamefont {Xiaoran}\
  \bibnamefont {Yan}}, \ and\ \bibinfo {author} {\bibfnamefont {Xin-Zeng}\
  \bibnamefont {Wu}},\ }\bibfield  {title} {\enquote {\bibinfo {title} {The"
  majority illusion" in social networks},}\ }\href@noop {} {\bibfield
  {journal} {\bibinfo  {journal} {PloS one}\ }\textbf {\bibinfo {volume}
  {11}},\ \bibinfo {pages} {e0147617} (\bibinfo {year} {2016})}\BibitemShut
  {NoStop}%
\bibitem [{\citenamefont {Eom}\ and\ \citenamefont
  {Jo}(2014)}]{eom2014generalized}%
  \BibitemOpen
  \bibfield  {author} {\bibinfo {author} {\bibfnamefont {Young-Ho}\
  \bibnamefont {Eom}}\ and\ \bibinfo {author} {\bibfnamefont {Hang-Hyun}\
  \bibnamefont {Jo}},\ }\bibfield  {title} {\enquote {\bibinfo {title}
  {Generalized friendship paradox in complex networks: The case of scientific
  collaboration},}\ }\href@noop {} {\bibfield  {journal} {\bibinfo  {journal}
  {Sci. Rep.}\ }\textbf {\bibinfo {volume} {4}},\ \bibinfo {pages} {srep04603}
  (\bibinfo {year} {2014})}\BibitemShut {NoStop}%
\bibitem [{\citenamefont {Yule}(1912)}]{yule1912on}%
  \BibitemOpen
  \bibfield  {author} {\bibinfo {author} {\bibfnamefont {George~Udny}\
  \bibnamefont {Yule}},\ }\bibfield  {title} {\enquote {\bibinfo {title} {On
  the methods of measuring association between two attributes},}\ }\href@noop
  {} {\bibfield  {journal} {\bibinfo  {journal} {J. R. Stat. Soc.}\ }\textbf
  {\bibinfo {volume} {75}},\ \bibinfo {pages} {579--652} (\bibinfo {year}
  {1912})}\BibitemShut {NoStop}%
\bibitem [{\citenamefont {Ferguson}(1941)}]{ferguson1941factorial}%
  \BibitemOpen
  \bibfield  {author} {\bibinfo {author} {\bibfnamefont {George~A}\
  \bibnamefont {Ferguson}},\ }\bibfield  {title} {\enquote {\bibinfo {title}
  {The factorial interpretation of test difficulty},}\ }\href@noop {}
  {\bibfield  {journal} {\bibinfo  {journal} {Psychometrika}\ }\textbf
  {\bibinfo {volume} {6}},\ \bibinfo {pages} {323--329} (\bibinfo {year}
  {1941})}\BibitemShut {NoStop}%
\bibitem [{\citenamefont {Guilford}(1950)}]{guilford1942fundamental}%
  \BibitemOpen
  \bibfield  {author} {\bibinfo {author} {\bibfnamefont {Joy~Paul}\
  \bibnamefont {Guilford}},\ }\href@noop {} {\emph {\bibinfo {title}
  {Fundamental statistics in psychology and education.}}}\ (\bibinfo
  {publisher} {McGraw-Hill},\ \bibinfo {year} {1950})\BibitemShut {NoStop}%
\bibitem [{\citenamefont {Davenport~Jr}\ and\ \citenamefont
  {El-Sanhurry}(1991)}]{davenport1991phi}%
  \BibitemOpen
  \bibfield  {author} {\bibinfo {author} {\bibfnamefont {Ernest~C}\
  \bibnamefont {Davenport~Jr}}\ and\ \bibinfo {author} {\bibfnamefont
  {Nader~A}\ \bibnamefont {El-Sanhurry}},\ }\bibfield  {title} {\enquote
  {\bibinfo {title} {Phi/phimax: review and synthesis},}\ }\href@noop {}
  {\bibfield  {journal} {\bibinfo  {journal} {Educ. Psychol. Meas.}\ }\textbf
  {\bibinfo {volume} {51}},\ \bibinfo {pages} {821--828} (\bibinfo {year}
  {1991})}\BibitemShut {NoStop}%
\bibitem [{\citenamefont {Cureton}(1959)}]{Cureton1959}%
  \BibitemOpen
  \bibfield  {author} {\bibinfo {author} {\bibfnamefont {Edward~E.}\
  \bibnamefont {Cureton}},\ }\bibfield  {title} {\enquote {\bibinfo {title}
  {Note on $\phi$/$\phi$max},}\ }\href@noop {} {\bibfield  {journal} {\bibinfo
  {journal} {Psychometrika}\ }\textbf {\bibinfo {volume} {24}},\ \bibinfo
  {pages} {89--91} (\bibinfo {year} {1959})}\BibitemShut {NoStop}%
\bibitem [{\citenamefont {Cohen}(1960)}]{cohen1960coefficient}%
  \BibitemOpen
  \bibfield  {author} {\bibinfo {author} {\bibfnamefont {Jacob}\ \bibnamefont
  {Cohen}},\ }\bibfield  {title} {\enquote {\bibinfo {title} {A coefficient of
  agreement for nominal scales},}\ }\href@noop {} {\bibfield  {journal}
  {\bibinfo  {journal} {Educ Psychol Meas}\ }\textbf {\bibinfo {volume} {20}},\
  \bibinfo {pages} {37--46} (\bibinfo {year} {1960})}\BibitemShut {NoStop}%
\bibitem [{\citenamefont {Boldi}\ \emph {et~al.}(2007)\citenamefont {Boldi},
  \citenamefont {Santini},\ and\ \citenamefont {Vigna}}]{boldi2007deeper}%
  \BibitemOpen
  \bibfield  {author} {\bibinfo {author} {\bibfnamefont {Paolo}\ \bibnamefont
  {Boldi}}, \bibinfo {author} {\bibfnamefont {Massimo}\ \bibnamefont
  {Santini}}, \ and\ \bibinfo {author} {\bibfnamefont {Sebastiano}\
  \bibnamefont {Vigna}},\ }\bibfield  {title} {\enquote {\bibinfo {title} {A
  deeper investigation of pagerank as a function of the damping factor},}\ }in\
  \href@noop {} {\emph {\bibinfo {booktitle} {Web Information Retrieval and
  Linear Algebra Algorithms}}},\ \bibinfo {series and number} {\bibinfo
  {number} {07071}}\ (\bibinfo {year} {2007})\BibitemShut {NoStop}%
\bibitem [{\citenamefont {Fosdick}\ \emph {et~al.}(2016)\citenamefont
  {Fosdick}, \citenamefont {Larremore}, \citenamefont {Nishimura},\ and\
  \citenamefont {Ugander}}]{fosdick2016configuring}%
  \BibitemOpen
  \bibfield  {author} {\bibinfo {author} {\bibfnamefont {Bailey~K}\
  \bibnamefont {Fosdick}}, \bibinfo {author} {\bibfnamefont {Daniel~B}\
  \bibnamefont {Larremore}}, \bibinfo {author} {\bibfnamefont {Joel}\
  \bibnamefont {Nishimura}}, \ and\ \bibinfo {author} {\bibfnamefont {Johan}\
  \bibnamefont {Ugander}},\ }\bibfield  {title} {\enquote {\bibinfo {title}
  {Configuring random graph models with fixed degree sequences},}\ }\href@noop
  {} {\bibfield  {journal} {\bibinfo  {journal} {preprint arXiv:1608.00607}\ }
  (\bibinfo {year} {2016})}\BibitemShut {NoStop}%
\end{thebibliography}%

\appendix

\section{Directed networks}
\label{app_dir}

We can easily extend the multiscale mixing measure $r_{\alpha}$ described in the main text to directed networks.  The main change is to incorporate two sets of marginals $a$ and $b$ that describe the proportion of edges starting from and ending at each of the attribute types.  Then the directed global assortativity of a network with respect to a particular categorical node attribute $y_i$ is:
\begin{equation}
  r_{\textrm{global}} = \frac{\sum_{g}e_{gg}- \sum_{g}a_gb_g}{1-\sum_{g}{a_gb_g}}  \enspace, \label{eq_assortativity_dir}
\end{equation}
where $a_g$ and $b_g$ represent the total number of outgoing and incoming links of all nodes of type $g$:
\begin{equation}
     a_{g} = \sum_{h} e_{gh}\enspace , \qquad b_{h} = \sum_{g} e_{gh} \enspace.
\end{equation}
Then we can update our definition of local assortativity accordingly,
\begin{equation}
  r(\ell) = \frac{1}{Q_{\rm{max}}} \sum_{g} (e_{gg}(\ell) - a_gb_g) \enspace.
\end{equation}

\section{Scalar attributes}
\label{app_sca}

\begin{figure*}
 \includegraphics[width=\linewidth]{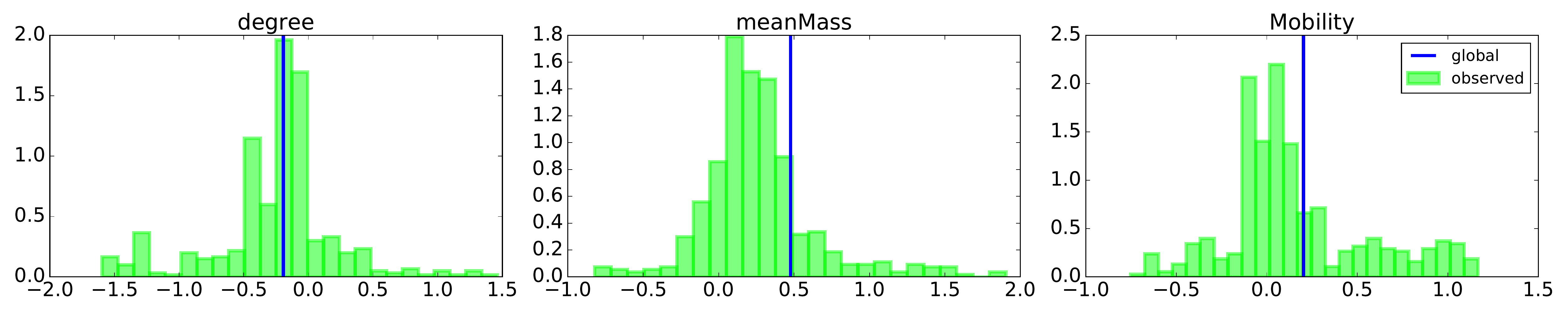}
 \caption{Multiscale assortativity for different scalar attributes in the Weddell Sea Food Web: node degree, average species mass and mobility.  Note that mobility is a discrete ordinal variable (taking integer values in [1,4]) and in the main text we treat it as an unordered discrete variable. }
 \label{fig_scalar}
\end{figure*}

For scalar attributes we can simply calculate the Pearson's correlation across edges.  Using $x_i$ and $x_j$ to indicate the scalar attribute value of the nodes in edge $A_{ij}$ then we can write the global assortativity as,
\begin{align}
  r_{\rm{global}} =& \frac{\textrm{cov}(x_i,x_j)}{\sigma_i \sigma_j} \\
  =& \frac{\sum_{ij}{A_{ij}(x_i - \bar{x})(x_j - \bar{x})}}
  { \sum_{i}{k_i(x_i - \bar{x})^2}} \enspace,
\end{align}
where $\bar{x} = \frac{1}{2m}\sum_i k_i x_i$ is the mean value of $x$ weighted by node degree $k$ and $\sigma_i$ is the standard deviation of the attribute values.  If we standardise the scalar values using the linear transformation 
$\tilde{x}_i = \frac{x_i - \bar{x}}{\sigma_i}$, then we can simplify this further as,
\begin{equation}
  r_{\rm{global}} = \sum_{ij} \frac{A_{ij}}{2m}\tilde{x}_i \tilde{x}_j \enspace .
  \label{eq:rglobal:scalar}
\end{equation}
Then we can calculate the local assortativity $r_{\alpha}(\ell)$ for scalar variables as,
\begin{equation}
  r_{\alpha}(\ell) = \sum_{ij} w_{\alpha}(i;\ell) \frac{A_{ij}}{k_i} \tilde{x}_i \tilde{x}_j \enspace.
\end{equation}
Figure~\ref{fig_scalar} gives some examples of distributions of $r_{\rm{multi}}$ for scalar attributes in the food web network.

\section{Categorical assortativity as a correlation}
\label{app_cat}
The assortativity coefficient $r_{\rm{global}}$ for categorical attributes  can be interpreted as a normalised Pearson's correlation.  To see this, we start by observing that the Pearson's correlation of two binary variables is equivalent to the Phi coefficient for binary contingency tables~\cite{yule1912on}. Table~\ref{tab_phi_contingency} shows a contingency table using the same notation as the directed assortativity, i.e., $a$ and $b$ give the marginal proportions and $e$ gives the joint proportions.
  
\begin{table}[h!]
 \caption{Binary contingency table}
 \label{tab_phi_contingency}
 \begin{center}
 \begin{tabular}{c|c|c|c} 
 &$y_j = 0$ & $y_j = 1$ & \\
 \hline
 $y_i = 0$&$e_{00}$ & $e_{01}$ & $a_{0}$ \\
 \hline
 $y_i = 1$&$e_{10}$ & $e_{11}$ & $a_{1}$ \\
 \hline
 &$b_{0}$ & $b_{1}$ &  \\
 \end{tabular}
 \end{center}
\end{table}

\noindent Then the Pearson product-moment correlation of these variables is known as $\phi$, which we derive using the moments of a Bernoulli distribution:
\begin{align}
 \phi = & \frac{\mathbb{E}[y_i,y_j] - \mathbb{E}[y_i]\mathbb{E}[y_j]}{\sigma_{y_i}\sigma_{y_j}} \\
 = &\frac{e_{11} - a_{1}b_{1}}{\sqrt{a_{1}a_{0}}\sqrt{b_{1}b_{0}}} \enspace .
\end{align}
Note that it is only necessary to calculate this in terms of $e_{11}$, since $e_{11} - a_{1}b_{1}=e_{00} - a_{0}b_{0}$.  We can see this using the identity $e_{00} = b_0 - a_1 +e_{11}$:
\begin{align}
  e_{00} - a_{0}b_{0} &= b_0 - a_1 +e_{11} - (1-a_{1})(1-b_{1})\\
  &= (1-b_{1})- a_1 +e_{11} - (1 - a_{1}-b_{1}+a_{1}b_{1}) \\
  &= e_{11} - a_{1}b_{1} \enspace .
\end{align}

A well-known issue with $\phi$ is that the extreme values of $+1$ and $-1$ are typically unobtainable, which can cause issues with its interpretation.  In fact $\phi=1$ can only occur if $a_{1}=b_{1}$, e.g., when the network is undirected, while $\phi=-1$ can only occur if $a_{1}=b_{2}=0.5$~\cite{ferguson1941factorial, guilford1942fundamental}.  To address this issue, there have been a number of proposed normalisations to ensure the $\phi=1$ is obtainable~\cite{davenport1991phi}.  One such normalisation is the $\phi/\phi_{\rm{max}}$ proposed by Cureton~\cite{Cureton1959},
\begin{equation}
\frac{\phi}{\phi_{\rm{max}}} = \frac{e_{11} - a_{1}b_{1}}{\beta - a_{1}b_{1} } \enspace ,
\end{equation}
where $\beta$ is the maximum possible value that $e_{11}$ can take, i.e., $\min (a_{1}b_{1})$. Note for undirected networks
\begin{align}
  \sqrt{a_{1}b_{1}a_{2}b_{2}} &= \sqrt{a_{1}^2a_{2}^2} \\
  &=a_{1}a_{2} \\
  &=a_{1}(1-a_{1}) \\
  &=a_{1} - a_{1}^2 \enspace ,
\end{align}
which equals $\phi_{\rm{max}}$ wnen $a_1 \leq a_2$.

 Then we can generalise $\phi/\phi_{\rm{max}}$ from binary to multi-category variables by treating each distinct value as a binary variable and taking their sum.  If we set $\beta=1$, then we obtain~\eqref{eq_assortativity_dir} and thus we recover Newman's assortativity~\cite{newman2003mixing}. We also note that~\eqref{eq_assortativity_dir} also corresponds to Cohen's $\kappa$ that is frequently used to assess inter-rater agreement~\cite{cohen1960coefficient}.

The normalisation of the assortativity coefficient means that $r_{\textrm{min}} \leq r \leq 1$ and 
\begin{equation}
  r_{\textrm{min}} = - \frac{\sum_{g}a_gb_g}{1-\sum_{g}{a_{g}b_{g}}}\enspace ,
\end{equation}
which lies in the range $-1 \leq r_{\textrm{min}} < 0$. 


\section{Assortativity as autocorrelation of a time-series}
\label{app_aut}

Assume a scalar attribute $x_i$ on each node $i$ of an undirected network. As mentioned in the main text, the probability of being at node $i$ is stationary and proportional to the degree, $\pi_i=k_i/2m$. Given that a random walker is currently at node $i$, it moves to node $j$ with probability $A_{ij}/k_i$.
 
We define a random time-series, using the simple random walker, as the sequence of attributes of the nodes visited in the random walk, i.e., the value of the time series at time $t$ is the attribute value $x$ of the node visited at time $t$ in the random walk. 
Asymptotically, the average value observed by the random walker is $\bar{x}=\sum_i \pi_i x_i= \sum k_i x_i /2m$ and
the variance is $\sigma^2=\sum_i \pi_i x_i^2 - \bar{x}^2$. 

Likewise, the autocovariance between the attribute observed at two consecutive steps (time lag of 1) is $R_x = \sum_{ij}  \pi_i \frac{A_{ij}}{k_i} x_i x_j  - \bar{x}^2$.
Replacing $x$ by $\tilde{x}=\frac{x-\bar{x}}{\sigma}$, we obtain the autocorrelation $R_{\tilde{x}}= \sum_{ij}  \pi_i \frac{A_{ij}}{k_i} \tilde{x}_i \tilde{x}_j$, which coincides with $r_{\rm{global}}$ as defined in \eqref{eq:rglobal:scalar}. 

When faced with categorical data, we proceed as in SI C. We consider for each type $g$ of nodes the scalar attribute $x_g$ valued at 1 for nodes with type $g$ and zero elsewhere. The modularity $Q$ is therefore the sum for each type $g$ of the autocovariance, $Q=\sum_g R_g$. As in SI C, this can be normalised in various ways, one of which is Newman's global assortativity as used in this article, which therefore represents a sort of categorical autocorrelation of the time-series process of the categorical attributes observed by the stationary simple random walker.

\section{Disconnected networks}
\label{app_dis}
By using the personalised PageRank as a neighbourhood function it means that only nodes within the same connected component contribute to $r_{\rm{multi}}$. Consequently $r_{\rm{multi}}$ for each node is insensitive to whether or not multiple connected components are included.

\section{Missing values}
\label{app_mis}
It is common when dealing with real datasets that some values may be missing.  This is the case for the Facebook100 data, where a number of node attributes are missing. When considering the global assortativity 
previous work has simply ignored contributions from missing data values~\cite{traud2012social}.  That is, only edges that connect nodes for which both the attribute values are known are considered when calculating $e_{gh}$. This treatment works fine for the global assortativity because each edge counts equally.  However, simply omitting missing values when calculating the local assortativity can cause a bias in the distribution.  For example, consider the case when node $\ell$ and its immediate neighbours have missing values, but beyond those the attribute values are known.  For small values of $\alpha$ the weight $w_{\alpha}(i;\ell)$ is largest for nodes with missing attribute values.  Simply ignoring their edges would mean reassigning more weight to edges further away from $\ell$ when normalising to ensure that $\sum_{gh} e_{gh}(\ell)=1$, a necessary step in calculating the assortativity. Then when we examine the distribution of $r(\ell)$ across all nodes in the network, the resulting distribution will be biased representation.  To deal with this issue we calculate each of the local assortativities as normal, but assign each a weight $z_{\ell} = \sum_{gh} e_{gh}(\ell)$, i.e., the sum of local edge counts before normalisation.  The weight $z_{\ell}$ describes our confidence in the local assortativity estimate from $z_{\ell}=0$, indicating no confidence, to $z_{\ell}=1$ when all node attributes within the neighbourhood are known.  We adjust for these weights when plotting the histograms in the main text.

\section{Calculating the personalised PageRank vector}
\label{app_ppr}
The personalised PageRank vector is the stationary distribution of a random walk with restarts. We calculate it by direct simulation of the random walk process using the power method:
\begin{equation}
  w_{\alpha}(i;\ell)_{s+1} = \alpha \sum_j \frac{A_{ij}}{k_i} w_{\alpha}(j;\ell)_{s} + (1-\alpha) \delta_{i,\ell} \enspace ,
  \label{eq_power}
\end{equation}
and at convergence yields a distribution $w(i;\ell)$ with a mode at $\ell$.

\section{Integrating over \texorpdfstring{$\alpha$}{a}}
\label{app_int}
To integrate over all values of $\alpha$, we take advantage of the fact that we can equivalently write the $\eta$-th approximation the power method in \eqref{eq_power} as the $\eta$-th degree truncation of the power series~\cite{boldi2007deeper}:
\begin{equation}
  w_{\alpha}(i;\ell)_{\eta} = \delta_{i,\ell} 
  + \sum_{s=1}^{\eta}{ \alpha^{s} \left[ \left(\frac{A_{i\ell}}{k_i}\right)^{s} - \left(\frac{A_{i\ell}}{k_i}\right)^{(s-1)} \right]} \enspace .
\end{equation}
By taking advantage of the relationship between $\alpha$ and the sequence of approximations computed by the power method, we can calculate the distribution $w_{\alpha}(i;\ell)$ for a given $\alpha=\alpha_0$ and use the sequence of approximations to calculate the distribution for any other $\alpha$~\cite{boldi2007deeper}:
\begin{equation}
  w_{\alpha}(i;\ell)_{\eta} = \delta_{i,\ell} + \sum_{s=1}^{\eta} \frac{\alpha^s}{\alpha_0^s} \left(w(i;\ell, \alpha_0)_{s} - w(i;\ell, \alpha_0)_{s-1}\right) \enspace .
\end{equation}
We can then integrate over all possible values of $\alpha$~\cite{boldi2005totalrank},
\begin{align}
 w_{\rm{multi}}(i;\ell)_{\eta} &= \int_{0}^{1} \! w_{\alpha}(i;\ell)_{\eta} \; \rm{d}\alpha \\
 &= 
 \delta_{i,\ell} + \sum_{s=1}^{\eta} \frac{\left(w_{\alpha_{0}}(i;\ell)_{s} - w_{\alpha_{0}}(i;\ell)_{s-1}\right)}{(s+1)\alpha_{0}^{s}}\enspace .
\end{align}

\section{Null model network generation}
\label{app_nul}
We created a null model to generate networks with the same global assortativity as the observed network to compare the distributions of $r_{\rm{multi}}$.  For a fair comparison, we decided to keep the node degree and metadata label fixed while randomly rewiring the network.  We do so using a modified version of the Markov chain Monte Carlo (MCMC) sampling of the configuration model for stub-labelled\footnote{for simple graphs sampling from the space of stub-labelled graphs is equivalent to sampling from the space of vertex-labelled graphs~\cite{fosdick2016configuring}} simple graphs~\cite{fosdick2016configuring}.  The modification is to ensure that we sample a graph with (approximately) the same global assortativity as the observed network.  We achieve this by adding a rejection sampling step based on the binomial likelihood of observing the number of edges between nodes of the same type $m_{\rm{in}}= m\sum_g e_{gg}$ given the proportion of edges required to maintain the global assortativity $\omega_{\rm{in}} = \sum_g e_{gg}$,
\begin{equation}
 L(G_i) = \log \binom{\vphantom{\sum} m}{ m_{\rm{in}}} (\omega_{\rm{in}})^{m_{\rm{in}}}(1 - \omega_{\rm{in}})^{m - m_{\rm{in}}} \enspace .
\end{equation}

The modified MCMC algorithm is shown in Algorithm~\ref{alg_stub}

\begin{algorithm}[H]
 \caption{stub-labeled MCMC \label{alg_stub} }
\begin{algorithmic}
\Require {initial simple graph $G_0$, initial temp $t_0$}
\Ensure {sequence of graphs $G_i$}
 \For {$i<$ number of graphs to sample} 
 \State choose two edges at random
 \State randomly choose one of the two possible swaps
 \If {edge swap would create a self-loop or multiedge} 
\State resample current graph: $G_{i} \leftarrow G_{i-1}$
\Else
\If {$Unif(0,1)<\exp \left(\frac{L(G_i) - L(G_{i-1})}{t_i} \right)$}
    \State swap the chosen edges, producing $G_{i}$
\Else
    \State reject $G_{i}$
\EndIf
\EndIf
\State $t_{i+1} \leftarrow \textrm{update} (t_i)$
\EndFor
\end{algorithmic}
\end{algorithm}

\section{Datasets}
\label{app_dat}

\subsection{Weddell Sea Food Web}
The food web of the Antarctic Weddell Sea~\cite{brose2005body} consists of 488 species and 15885 consumer relations.  For each of the nodes in this network we have five categorical attributes: Metabolic Category \{\textit{Plant, Ectotherm vertebrate, Endotherm vertebrate, Invertebrate}\}, Feeding Type \{\textit{Carnivorous/necrovorous, Herbivorous/detrivorous, Detrivorous, Omnivorous, Primary producer, Carnivorous}\}, FeedingMode \{\textit{Pelagic predator, Predator/scavenger, Primary producer, Predator, Deposit-feeder, Grazer, Suspension-feeder}\}, Mobility $\{1, 2, 3, 4\}$, Environment \{\textit{Bathydemersal, Land-based, Resource, Pelagic, Benthopelagic, Benthic, Demersal}\}.  For scalar attributes we use the mean mass of the species, mobility (although discrete, the values are ordinal), and node degree.

\subsection{Facebook 100}The Facebook100 dataset~\cite{traud2012social} contains an anonymised snapshot of the friendship connections among $1,208,316$ users affiliated with
the first 100 colleges admitted to Facebook. The dataset contains a total of $93,969,074$ friendship edges between users of the same college. Each node has a set of categorical social variables: status \{\textit{undergraduate, graduate student, summer student, faculty, staff, alumni}\}, dorm, major, gender \{\textit{male, female}\}, and graduation year.

\begin{table}
\setlength{\tabcolsep}{5pt}
\renewcommand{\arraystretch}{1}
 \caption{List of schools ordered by global assortativity}
 \label{tab_school_assort}
 \begin{tabular}{|r|l|r|r|l|r|} 
 \hline
\# & School & $r_{\rm{glob}}$ & \# & School & $r_{\rm{glob}}$ \\
\hline
1  &  Amherst 41   &  0.081  &  51  &  William 77   &  0.203 \\
2  &  Princeton 12   &  0.087  &  52  &  Emory 27   &  0.205 \\
3  &  Trinity 100   &  0.106  &  53  &  UCLA 26   &  0.208 \\
4  &  Stanford 3   &  0.109  &  54  &  Tennessee 95   &  0.209 \\
5  &  Swarthmore 42   &  0.109  &  55  &  Wake 73   &  0.212 \\
6  &  Johns Hopkins 55   &  0.110  &  56  &  MIT 8   &  0.219 \\
7  &  Hamilton 46   &  0.113  &  57  &  UMass 92   &  0.222 \\
8  &  Bowdoin 47   &  0.118  &  58  &  Berkeley 13   &  0.222 \\
9  &  Harvard 1   &  0.120  &  59  &  USC 35   &  0.224 \\
10  &  Brown 11   &  0.120  &  60  &  Temple 83   &  0.228 \\
11  &  Dartmouth 6   &  0.126  &  61  &  UVA 16   &  0.230 \\
12  &  Wellesley 22   &  0.127  &  62  &  Penn 94   &  0.231 \\
13  &  Haverford 76   &  0.128  &  63  &  Northwestern 25   &  0.234 \\
14  &  Wesleyan 43   &  0.128  &  64  &  Rutgers 89   &  0.235 \\
15  &  UConn 91   &  0.129  &  65  &  UPenn 7   &  0.235 \\
16  &  Tufts 18   &  0.130  &  66  &  Michigan 23   &  0.236 \\
17  &  Williams 40   &  0.133  &  67  &  FSU 53   &  0.238 \\
18  &  Reed 98   &  0.134  &  68  &  Cornell 5   &  0.238 \\
19  &  Columbia 2   &  0.136  &  69  &  UC 64   &  0.251 \\
20  &  BC 17   &  0.136  &  70  &  American 75   &  0.253 \\
21  &  Duke 14   &  0.144  &  71  &  Notre Dame 57   &  0.255 \\
22  &  Virginia 63   &  0.149  &  72  &  Rochester 38   &  0.256 \\
23  &  Oberlin 44   &  0.151  &  73  &  Vassar 85   &  0.256 \\
24  &  Villanova 62   &  0.158  &  74  &  Lehigh 96   &  0.258 \\
25  &  Howard 90   &  0.159  &  75  &  Texas 80   &  0.261 \\
26  &  WashU 32   &  0.162  &  76  &  USFCA 72   &  0.263 \\
27  &  Georgetown 15   &  0.162  &  77  &  UC 61   &  0.265 \\
28  &  Colgate 88   &  0.164  &  78  &  Syracuse 56   &  0.270 \\
29  &  UF 21   &  0.165  &  79  &  Yale 4   &  0.273 \\
30  &  BU 10   &  0.167  &  80  &  UCSB 37   &  0.277 \\
31  &  Carnegie 49   &  0.171  &  81  &  Cal 65   &  0.279 \\
32  &  GWU 54   &  0.171  &  82  &  Texas 84   &  0.291 \\
33  &  Bingham 82   &  0.176  &  83  &  UChicago 30   &  0.291 \\
34  &  NYU 9   &  0.182  &  84  &  Smith 60   &  0.292 \\
35  &  UNC 28   &  0.185  &  85  &  Mississippi 66   &  0.297 \\
36  &  Simmons 81   &  0.186  &  86  &  Baylor 93   &  0.297 \\
37  &  USF 51   &  0.187  &  87  &  UIllinios 20   &  0.297 \\
38  &  JMU 79   &  0.187  &  88  &  MU 78   &  0.306 \\
39  &  UCF 52   &  0.187  &  89  &  Tulane 29   &  0.313 \\
40  &  Santa 74   &  0.188  &  90  &  Mich 67   &  0.322 \\
41  &  Northeastern 19   &  0.190  &  91  &  UGA 50   &  0.336 \\
42  &  Maine 59   &  0.190  &  92  &  Wisconsin 87   &  0.338 \\
43  &  Middlebury 45   &  0.190  &  93  &  UCSD 34   &  0.355 \\
44  &  Brandeis 99   &  0.193  &  94  &  Indiana 69   &  0.356 \\
45  &  Bucknell 39   &  0.194  &  95  &  UC 33   &  0.361 \\
46  &  MSU 24   &  0.195  &  96  &  Auburn 71   &  0.370 \\
47  &  Pepperdine 86   &  0.198  &  97  &  Oklahoma 97   &  0.397 \\
48  &  Vermont 70   &  0.199  &  98  &  Caltech 36   &  0.426 \\
49  &  Maryland 58   &  0.199  &  99  &  UCSC 68   &  0.480 \\
50  &  Vanderbilt 48   &  0.201  &  100  &  Rice 31   &  0.504 \\
 \hline
 \end{tabular}
\end{table}

\begin{figure*}
	\includegraphics[width=\linewidth]{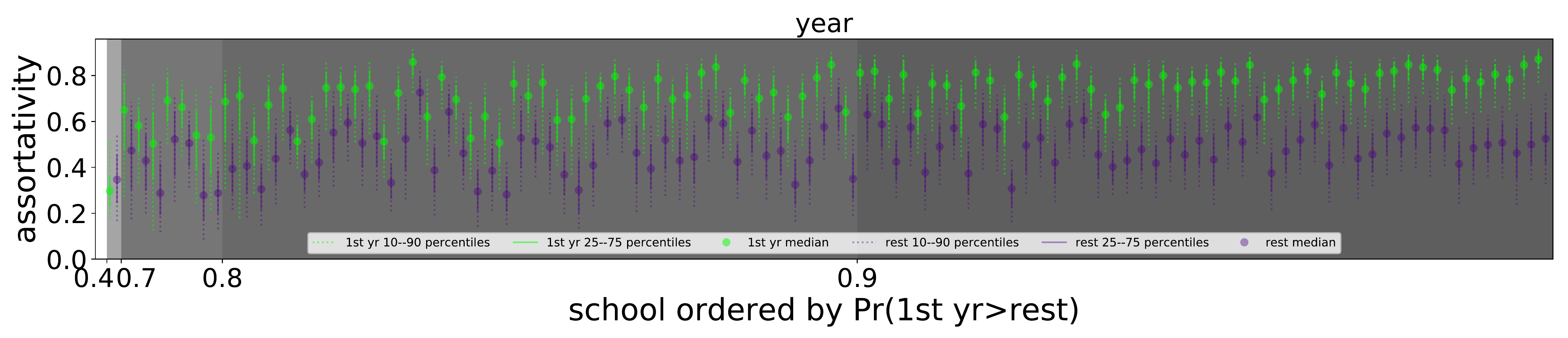}
	\caption{Distributions of the local assortativity by year separated into first years and the rest of the students for each school.  Schools are ordered by increasing proportion of first years that are more assortative than the rest of the students.  First year students are more assortative than the rest for all schools except for one.}
\label{fig_fb_year}
\end{figure*}

\begin{figure*}
	\includegraphics[width=\linewidth]{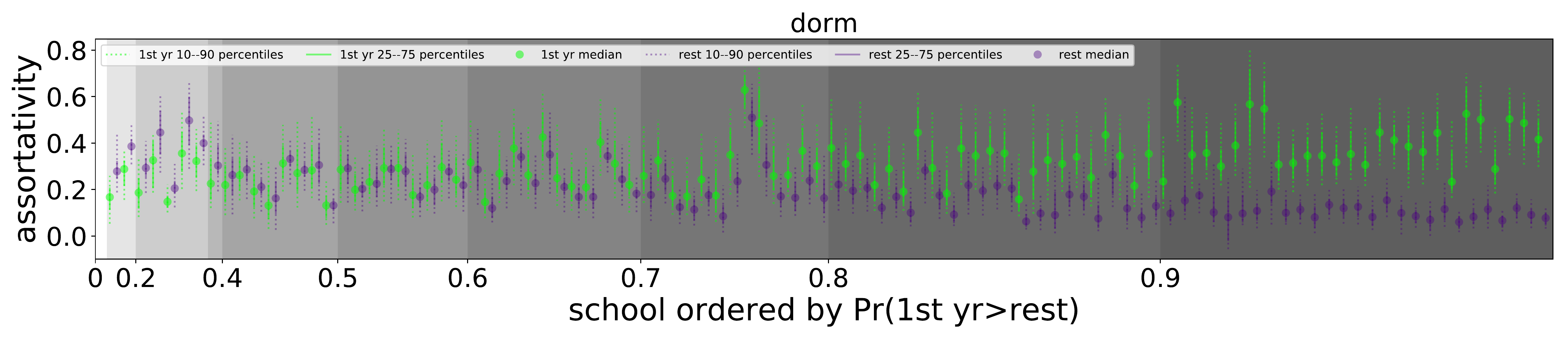}
	\caption{Distributions of the local assortativity by residence (dorm) separated into first years and the rest of the students for each school.  Schools are ordered by increasing proportion of first years that are more assortative than the rest of the students.  In general, first year students are more assortative than the rest, however there are some schools in which the difference between first year and the rest is negligible. In a few schools we observe that the first year students are less assortative than the rest.}
\label{fig_fb_dorm}
\end{figure*}

\clearpage

\end{document}